\documentstyle[aps,preprint,floats,psfig]{revtex}

\begin{document}
\newcommand{\beq}{\begin{equation}}
\newcommand{\eeq}{\end{equation}}
\draft
\title  {Tunneling current in triplet $f$-wave superconductors with horizontal
	lines of nodes}

\author{N. Stefanakis}
\address{ Department of Physics, University of Crete,
	P.O. Box 2208, GR-71003, Heraklion, Crete, Greece}

\date{\today}
 
\maketitle
\begin{abstract}

We calculate the tunneling conductance spectra of a 
normal-metal/insulator/triplet superconductor 
using the 
Blonder-Tinkham-Klapwijk (BTK) formulation. 
Possible states for the superconductor are 
considered with horizontal lines of nodes, breaking 
the time reversal symmetry. 
These results would be useful to discriminate between pairing states
in superonductor Sr$_2$RuO$_4$ and also in UPt$_3$.
\end{abstract}



\noindent{\it 1. Introduction}

The resent discovery of superconductivity in Sr$_2$RuO$_4$ 
has attracted much theoretical and experimental interest \cite{maeno}.
Knight-shift measurements show no change when passing 
through the superconducting state and is a clear 
evidence for spin triplet pairing state \cite{ishida}.
Muon spin rotation experiments show that the time reversal 
symmetry is broken for the superconductor Sr$_2$RuO$_4$ \cite{luke1}.
The linear temperature dependence of the nuclear spin 
lattice relaxation rate $1/T_1$ of $^{101}$Ru bellow $0.4K$ 
\cite{ishida1}, and specific heat measurements \cite{nishizaki}
are consistent with the presence of 
line nodes within the 
gap as in the high $T_c$ cuprate superconductors.

Knight-shift measurements show
that the parity of the pairing function of UPt$_3$
is odd and a spin triplet pairing state is
realized \cite{tou}.
Muon spin rotation experiments show that the time reversal
symmetry is broken bellow $T_{c2}$ for the superconductor UPt$_3$ \cite{luke}.
The nuclear spin
lattice relaxation rate $1/T_1$
and specific heat measurements
are consistent with the presence of
line nodes within the
gap as in the high $T_c$ cuprate superconductors.

In tunneling experiments involving singlet superconductors 
both line nodes and time reversal 
symmetry breaking can be detected from the V-like shape of the 
spectra, and the splitting of the zero energy conductance peak (ZEP)
at low temperatures respectively 
\cite{blonder,andreev,covington,stefan}.
Electron tunneling in Sr$_2$RuO$_4$ has been studied in Refs. 
\cite{stefan1,stefan2,yamashiro1}
for spin triplet pairing states with vertical lines of nodes. 
Electron tunneling in UPt$_3$ has been studied in Ref. \cite{yamashiro2}.

In this paper we discuss the tunneling effect in 
normal-metal/triplet superconductor with horizontal line nodes 
taking into account three dimensional effects.
For the triplet superconductor Sr$_2$RuO$_4$ we shall assume 
three possible pairing states of three
dimensional order parameter,
having horizontal lines of nodes, which run parallel to the basal plane and
break the time reversal symmetry.
The first two are the $f$-wave states 
proposed by Hasegawa {\it et al.},
\cite{hasegawa} having $A_{1g}\times E_u$ symmetry.
The other one is the $f$-wave pairing symmetry
proposed by H. Won and K. Maki \cite{won}.

For the triplet superconductor UPt$_3$ we shall assume
two possible pairing states of three
dimensional order parameter,
having horizontal lines of nodes, which run parallel to the basal plane and
break the time reversal symmetry.
These are the planar and bipolar pairing states
proposed by Machida {\it et al.} \cite{machida}.

\noindent{\it 2. Theory for the tunneling conductance}

\begin{figure}
  \centerline{\psfig{figure=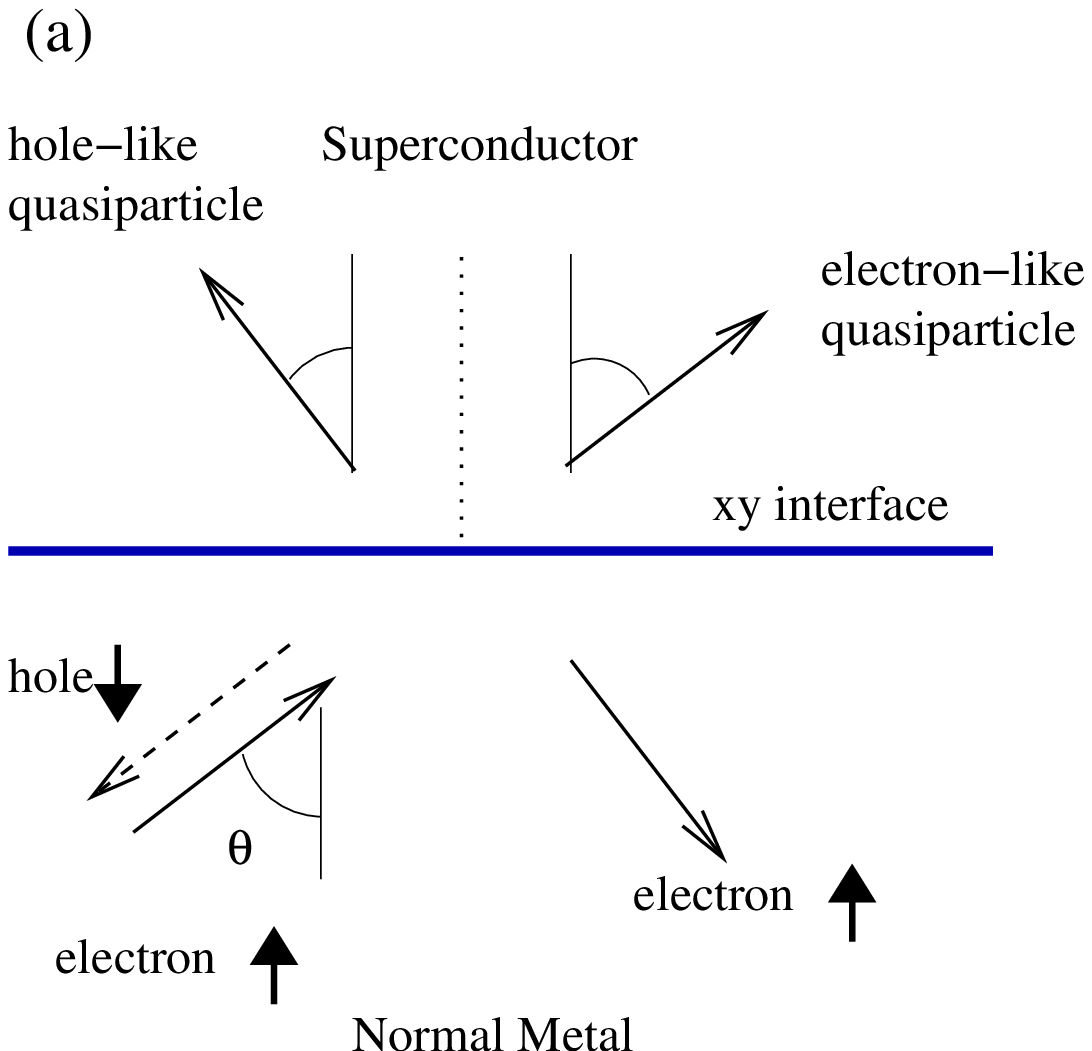,width=6.5cm,angle=0}}
  \centerline{\psfig{figure=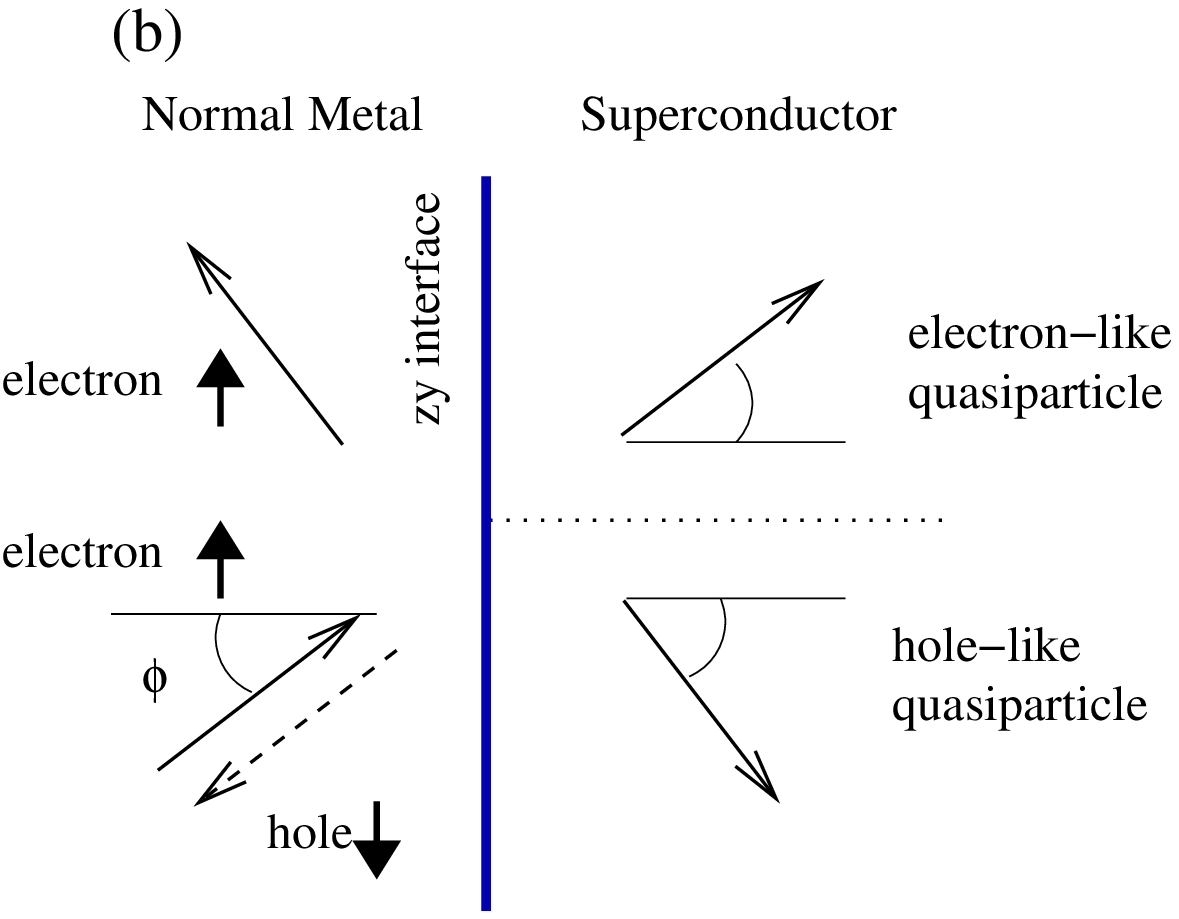,width=6.5cm,angle=0}}
  \caption{
The figure illustrate the transmission and reflection processes 
of the quasiparticle at the
interface of the junction 
with (a) $xy$ plane interface, (b) $zy$ plane interface.
}
  \label{interface.fig}
\end{figure}

The interface has a $\delta$-functional form 
is perpendicular to the $z$-axis and is located at $z=0$ as 
seen in Fig. \ref{interface.fig}(a) ($xy$-interface). 
Alternatively we consider the 
situation where the interface is perpendicular to the $x$-axis 
and is located at $x=0$ ($zy$-interface), (see Fig. \ref{interface.fig}(b)).
We assume a semi-infinite double layer structure and a spherical
Fermi surface. The motion of quasiparticles in inhomogeneous 
superconductors is described by the solution of the 
Bogoliubov-deGennes(BdG) equations. The effective pair potential 
is given by
\begin{equation} 
\Delta_{\rho \rho^{'}}({\bf k,r})=\Delta_{\rho \rho^{'}}(\phi,\theta)
\Theta(z)[\Theta(x)],
\end{equation} 
for the $xy$[$zy$]-interface, where 
$k_x,k_y,k_z=\cos\phi \sin\theta,\sin\phi \sin\theta,\cos\theta$.
$\phi$ is the azimuthal angle in the $xy$-plane, $\theta$ is the polar angle.
The quantities 
$\rho, \rho^{'}$ denote spin indices.
 
Suppose that an electron is injected from the normal metal with  
momentum $k_x,k_y,k_z$, and the interface is perpendicular to the 
$z$-axis. The electron
(hole) like quasiparticle will experience different pair potentials 
$\Delta_{\rho \rho^{'}}(\phi,\theta)$ 
($\Delta_{\rho \rho^{'}}(\phi,\pi-\theta)$). When the interface is 
perpendicular to the $x$-axis, the electron
(hole) like quasiparticle will experience different pair potentials
$\Delta_{\rho \rho^{'}}(\phi,\theta)$ 
($\Delta_{\rho \rho^{'}}(\pi-\phi,\theta)$).
The coefficients of the Andreev and normal reflection for the $xy$-interface 
are obtained by solving 
the BdG equations under the following boundary conditions 
\begin{equation}
\Psi({\bf r})|_{z=0_{-}}=\Psi({\bf r})|_{z=0_{+}}
\end{equation}
\begin{equation}
\frac{d\Psi({\bf r})}{dz}|_{z=0_{-}}=\frac{d\Psi({\bf r})}{dz}|_{z=0_{+}}-
\frac{2mV}{\hbar^2}\Psi({\bf r})|_{z=0_{-}}
\end{equation}
while for the $zy$ interface the boundary conditions are
\begin{equation}
\Psi({\bf r})|_{x=0_{-}}=\Psi({\bf r})|_{x=0_{+}}
\end{equation}
\begin{equation}
\frac{d\Psi({\bf r})}{dx}|_{x=0_{-}}=\frac{d\Psi({\bf r})}{dx}|_{x=0_{+}}-
\frac{2mV}{\hbar^2}\Psi({\bf r})|_{x=0_{-}}
\end{equation}

Using the obtained coefficients the tunneling conductance for the $xy$
interface is calculated using the formula 
$\sigma_(E)=\sigma_{\uparrow}(E)+\sigma_{\downarrow}(E)$, where the conductance 
for spin-up(-down) quasiparticle is given by the relation
\begin{equation}
\sigma_{\uparrow [\downarrow]}(E)=
	\frac{
\int_{0}^{2\pi}\int_{0}^{\frac{\pi}{2}} 
\overline{\sigma}_{\uparrow [\downarrow]}(E,\phi,\theta)
\sin \theta \cos \theta d\phi d\theta}
{\int_{0}^{2\pi}\int_{0}^{\frac{\pi}{2}} 
2\sigma_N\sin \theta \cos \theta d\phi d\theta}
,\label{xysE}
\end{equation}
where the normal-state conductance is given by
\begin{equation}
\sigma_N=\frac{\cos^2\theta}{\cos^2\theta+z_0^2}
.~~~\label{xysN}
\end{equation}

The corresponding formula for the tunneling conductance for the $zy$
interface is 
\begin{equation}
\sigma_{\uparrow [\downarrow]}(E)=
	\frac{
\int_{-\pi/2}^{\pi/2}\int_{0}^{\frac{\pi}{2}}  
\overline{\sigma}_{\uparrow [\downarrow]}(E,\phi,\theta)
	\sin^2 \theta \cos \phi d\phi d\theta}
{\int_{-\pi/2}^{\pi/2}\int_{0}^{\frac{\pi}{2}}
 \sin^2 \theta \cos \phi 
2\sigma_Nd\phi d \theta}
,\label{zysE}
\end{equation}
where the normal-state conductance is given by
\begin{equation}
\sigma_N=\frac{\cos^2\phi \sin^2 \theta}{\cos^2\phi \sin^2 \theta+z_0^2}
.~~~\label{zysN}
\end{equation}

The pairing potential is described by a $2\times 2$ form
\begin{equation}
\hat{\Delta}_{\alpha,\beta}({\bf k}) =
  \left(
    \begin{array}{ll}
     -d_x({\bf k}) +id_y({\bf k})&
     d_z({\bf k}) \\
     d_z({\bf k}) &
    d_x({\bf k}) +id_y({\bf k})
    \end{array}
  \right)
,~~~\label{deltafour}
\end{equation}
in terms of the $d({\bf k})=(d_x({\bf k}),d_y({\bf k}),d_z({\bf k}))$ vector.

For Sr$_2$RuO$_4$ the $d$-vector runs parallel to $z$-axis (i.e., 
$d({\bf k})=(0,0,d_z({\bf k})$). The candidate pairing states are given by

a) $d_z({\bf k})=(k_x+ik_y)\cos(ck_z)$,
with $c$ being the lattice constant along the 
$c$-axis. This state has horizontal lines of nodes at $k_z=\pm \frac{\pi}{2c}$
and breaks the time reversal symmetry. 

b) $d_z({\bf k})=(\sin(\frac{ak_x}{2})+i\sin(\frac{ak_y}{2}))\cos(\frac{ck_z}{2})$, 
with horizontal lines of nodes at $k_z=\pm \frac{\pi}{c}$.

c) $d_z({\bf k})=(k_x+ik_y)^2k_z$,
with horizontal lines of nodes at $k_z=0$.

Then we will choose two candidate pairing states corresponding to the
$B$-phase of UPt$_3$ (low temperature $T$ and low field $H$):
a) The unitary planar state with
$d({\bf k})=(\lambda_x({\bf k}),\lambda_y({\bf k}),0)$,
and b) the non-unitary bipolar state with
$d({\bf k})=(\lambda_x({\bf k}),i\lambda_y({\bf k}),0)$, where
$\lambda_x({\bf k})=k_z(k_x^2-k_y^2)$,
$\lambda_y({\bf k})=k_z2k_xk_y$.

According to the BTK formula the conductance 
of the junction, 
$\overline{\sigma}_{\uparrow [\downarrow]}(E)$, 
for up(down) spin quasiparticles, 
is expressed in terms of the 
probability amplitudes
$a_{\uparrow [\downarrow]},b_{\uparrow [\downarrow]}$ as
\cite{blonder}
\begin{equation}
\overline{\sigma}_{\uparrow [\downarrow]}(E) 
=1+|a_{\uparrow [\downarrow]}|^2
-|b_{\uparrow [\downarrow]}|^2
.~~~\label{ovs}
\end{equation}

The Andreev and normal reflection amplitudes
$a_{\uparrow [\downarrow]},b_{\uparrow [\downarrow]}$
for the spin-up(-down) quasiparticles are obtained as

\begin{equation}
a_{\uparrow [\downarrow]}=\frac{4n_{+}}
     {4+z_0^2
     -z_0^2n_{+}n_{-}
      \phi_{-}\phi_{+}^{\ast}}
,~~~\label{ra}
\end{equation}

\begin{equation}
b_{\uparrow [\downarrow]}=\frac
     {-(2iz_0+z_0^2)
	+(2iz_0+z_0^2)n_{+}n_{-}
      \phi_{-}\phi_{+}^{\ast}}
{4+z_0^2-z_0^2n_{+}n_{-}\phi_{-}\phi_{+}^{\ast}}
,~~~\label{rb}
\end{equation}
where 
$z_0=\frac{m V}{\hbar^2 k_s}$. 
The BCS coherence factors are given by 
\begin{equation}
u_{\pm}^2=[1+
      \sqrt{E^2-|\Delta_{\pm}|^2}/E]/2,
\end{equation}
\begin{equation}
v_{\pm}^2=[1-
      \sqrt{E^2-|\Delta_{\pm}|^2}/E]/2,
\end{equation}
and $n_{\pm}=v_{\pm}/u_{\pm}$.
The internal phase coming from the energy gap is given by
$\phi_{\pm} =[
\Delta_{\pm}/|\Delta_{\pm}|]$,
where $\Delta_{+}$
($\Delta_{-}$) is the 
pair potential experienced by the transmitted electron-like 
(hole-like) quasiparticle.

\noindent{\it 3. Sr$_2$RuO$_4$}

\begin{figure}
 \centerline{\psfig{figure=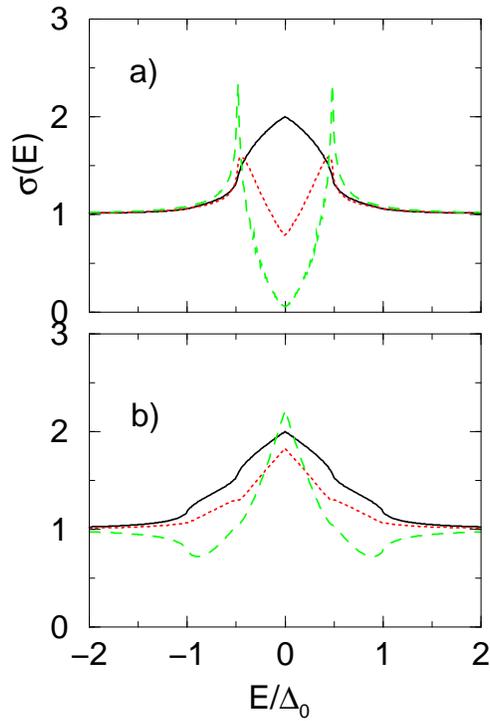,width=6.5cm,angle=0}}
  \caption{
Normalized tunneling conductance $\sigma(E)$ as a function of $E/\Delta_0$
for $z_0=0$ (solid line), $z_0=0.5$ (dotted line), $z_0=2.5$ (dashed line), 
for the superconductor Sr$_2$RuO$_4$. 
In (a) the interface is perpendicular to the $z$-axis, and in (b)
the interface is perpendicular to the $x$-axis.
The pairing 
symmetry of the superconductor is 
$(k_x+ik_y)\cos(ck_z)$-wave.
}
  \label{Eucosckz.fig}
\end{figure}

\begin{figure}
 \centerline{\psfig{figure=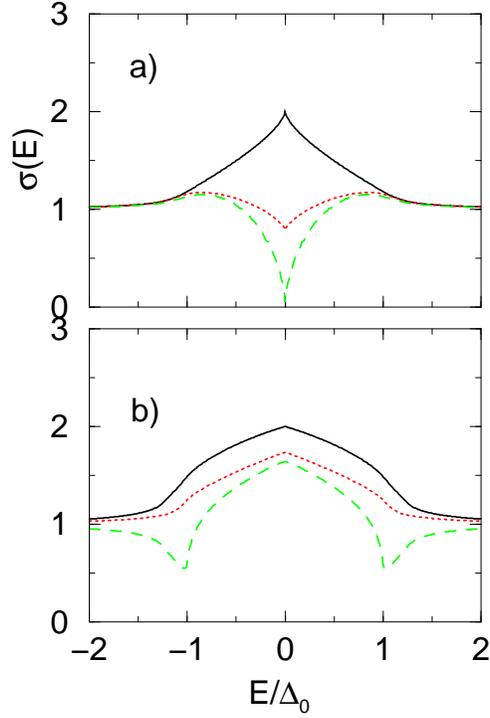,width=6.5cm,angle=0}}
  \caption{
The same as in Fig. 3. 
The pairing 
symmetry of the superconductor is 
$(\sin(\frac{ak_x}{2})+i\sin(\frac{ak_y}{2}))\cos(\frac{ck_z}{2})$-wave.
}
  \label{sincos.fig}
\end{figure}

\begin{figure}
 \centerline{\psfig{figure=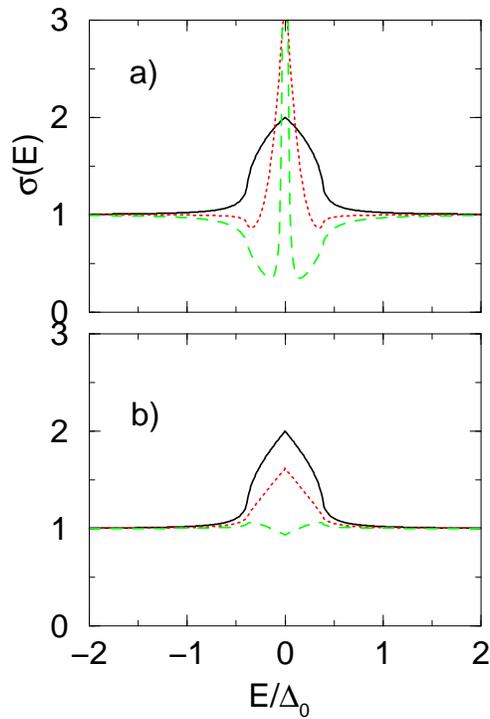,width=6.5cm,angle=0}}
  \caption{The same as in Fig. 3.
The pairing
symmetry of the superconductor is $(k_x+ik_y)^2k_z$-wave.}
  \label{Eu2kz.fig}
\end{figure}

In Figs. \ref{Eucosckz.fig}-\ref{Eu2kz.fig} 
we plot the tunneling conductance $\sigma(E)$
as a function of $E/ \Delta_0$
for various values of $z_0$, for the $xy$-interface (a), and 
$zy$-interface (b), for the superconductor Sr$_2$RuO$_4$.
The pairing symmetry of the superconductor is
$(k_x+ik_y)\cos(ck_z)$-wave in Fig. \ref{Eucosckz.fig},
$(\sin(\frac{ak_x}{2})+i\sin(\frac{ak_y}{2}))\cos(\frac{ck_z}{2})$-wave 
in Fig. \ref{sincos.fig},
$(k_x+ik_y)^2k_z$-wave in Fig. \ref{Eu2kz.fig}.

The conductance peak is formed
in the electron tunneling for the $xy$-interface,  
$zy$-interface when
the transmitted quasiparticles experience different sign of the
pair potential on the Fermi surface (FS). 
Also the line shape of the spectra is sensitive
to the presence or absence of nodes of the pair potential on the
Fermi surface.

For the $(k_x+ik_y)\cos(ck_z)$-wave case, for the $xy$-interface
the scattering process changes the electron momentum from 
$(\phi,\pi-\theta)$ to $(\phi,\theta)$ on the 
FS. However this process conserves the sign of the pair 
potential for $0 < \phi < 2\pi$. 
As a result no peak exists in the conductance spectra, 
as seen in Fig. \ref{Eucosckz.fig} (a) for $z_0=2.5$.
Also the nodes of the pair potential at $k_z=\pm \pi/2c$
intersect the FS along the $z$-axis
and a V-like gap opens in the tunneling spectra
as in the case of the $d$-wave superconductor.
On the other hand for the $zy$-interface
the transmitted quasiparticles feel different sign of the 
pair potential for $(\phi,\theta)$ and $(\pi-\phi,\theta)$
only 
at discrete $\phi$-values which explain the residual values 
of the conductance within the energy gap seen in 
Fig. \ref{Eucosckz.fig} (b). 

For the $(\sin(\frac{ak_x}{2})+i\sin(\frac{ak_y}{2}))\cos(\frac{ck_z}{2})$-wave
case, and for the $xy$-interface 
the scattering process in the momentum space
connects points of the
FS with the same sign as in the $(k_x+ik_y)\cos(ck_z)$-wave case.
This means that the pair potential does not change 
sign and no ZEP is formed as seen in Fig. \ref{sincos.fig} (a). 
However the pair potential 
intersect the spherical FS at the poles, i.e., at $k_z=\pm \pi/c$, 
forming point like nodes. This explains the logarithmic singularity 
at $E=0$ at the spectra.
The tunneling spectra for the
$zy$-interface is enhanced due to the bound states 
that are formed at discrete values 
of the quasiparticle angle $\phi$ as seen in Fig. \ref{sincos.fig} (b).
 
The situation is opposite in the $(k_x+ik_y)^2k_z$-wave case where
the scattering process for $xy$ interface
connects points of the FS, i.e., $(\phi,\pi-\theta)$ and $(\phi,\theta)$, with opposite sign.
As a consequence the ZEP is formed, for $0 < \phi < 2\pi$
as seen in Fig. \ref{Eu2kz.fig} (a). 
Also in this case the node of the pair potential at $k_z=0$ 
intersect the FS and 
the spectra has a V-shaped form as in the $(k_x+ik_y)\cos(ck_z)$-wave case. 
For the $zy$-interface the order parameter has the same $\phi$
dependence as in the $(k_x+ik_y)\cos(ck_z)$-wave, 
and the tunneling spectra for the 
$zy$-interface is similar.

The conclusion is that the tunneling at the $xy$ interface can be used 
to distinguish the pairing states with horizontal lines of nodes 
on the FS. The numerical results presented here are in agreement with 
recent analytical calculation for the case of the low transparency barrier 
and triplet pairing states with horizontal lines of nodes \cite{sengupta}.
Also only one electron band for Sr$_2$RuO$_4$ contributes to superconductivity.
However it has been proposed recently that the $\gamma$ band is nodeless 
while $\alpha$ and $\beta$ bands 
have horizontal lines of nodes \cite{zhitomirsky}. 
In this case all three bands contribute to the pairing state and the 
actual shape 
of the spectra depends on the amount of contribution of different bands. 

In this paragraph a comparison is made between the pairing states examined 
here and the corresponding states without the $k_z$-dependence.
For the $k_x+ik_y$-wave case, for the $xy$ interface the gap has a 
U-shaped structure due to the absence of sign change of the order 
parameter along the $z$-axis. For the $zy$-interface the difference 
is more pronounced at $z_0=0$ where $\sigma(E)$ is constant within the 
gap for the $k_x+ik_y$-wave case, while it posses a $\Lambda$ shape 
structure for the $(k_x+ik_y)\cos(ck_z)$-wave.

For the $\sin(\frac{ak_x}{2})+i\sin(\frac{ak_y}{2})$-wave, and $xy$-interface
the spectra is expected to have the U-shaped line shape, while for the 
$zy$ interface the spectra should have residual values due to the 
presence of bound states. However a detailed calculation is needed 
to account for the actual shape of the spectra.

\noindent{\it 4. UPt$_3$}

\begin{figure}
 \centerline{\psfig{figure=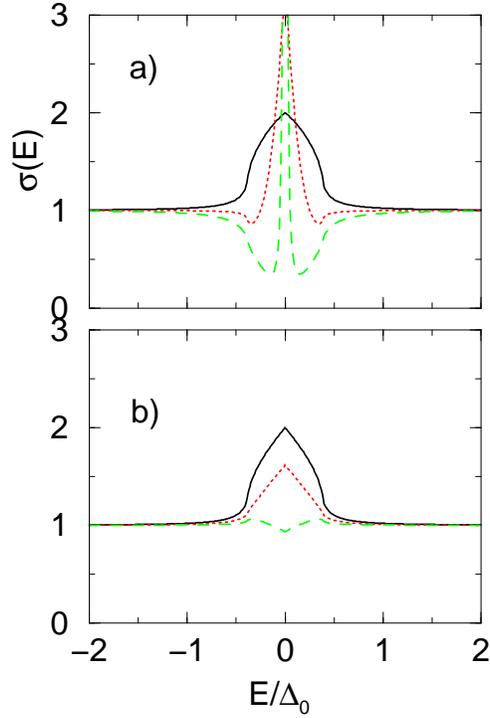,width=6.5cm,angle=0}}
  \caption{
Normalized tunneling conductance $\sigma(E)$ as a function of $E/\Delta_0$
for $z_0=0$ (solid line), $z_0=0.5$ (dotted line), $z_0=2.5$ (dashed line),
for the superconductor UPt$_3$.
The pairing
symmetry of the superconductor is
planar.
}
  \label{planar.fig}
\end{figure}

\begin{figure}
 \centerline{\psfig{figure=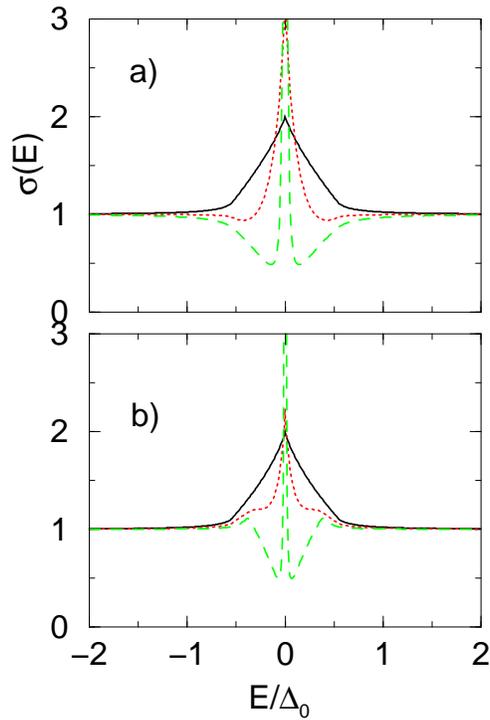,width=6.5cm,angle=0}}
  \caption{
The same as in Fig. 3.
The pairing
symmetry of the superconductor is
bipolar.
}
  \label{bipolar.fig}
\end{figure}

In Figs. \ref{planar.fig}-\ref{bipolar.fig}
we plot the tunneling conductance $\sigma(E)$
as a function of $E/ \Delta_0$
for various values of $z_0$, for the $xy$-interface (a), and
$zy$-interface (b) for the superconductor UPt$_3$.
The pairing symmetry of the superconductor is
the unitary planar in Fig. \ref{planar.fig} and the 
non unitary bipolar
in Fig. \ref{bipolar.fig}.

The conductance peak is formed
in the electron tunneling for the $xy$-interface,
$zy$-interface when
the transmitted quasiparticles experience different sign of the
pair potential on the Fermi surface.
Also the line shape of the spectra is sensitive
to the presence or absence of nodes of the pair potential on the
Fermi surface.

For the planar case, for the $xy$-interface
the scattering process changes the electron momentum from
$(\phi,\pi-\theta)$ to $(\phi,\theta)$ on the
FS. This process changes the sign of the pair
potential for $0 < \phi < 2\pi$.
As a result a peak exists in the conductance spectra,
as seen in Fig. \ref{planar.fig} (a) for $z_0=2.5$.
Also the nodes of the pair potential at $k_z=0$
intersect the FS along the $z$-axis
and a V-like gap opens in the tunneling spectra
as in the case of the $d$-wave superconductor.
On the other hand for the $zy$-interface
the transmitted quasiparticles feel different sign of the
pair potential for $(\phi,\theta)$ and $(\pi-\phi,\theta)$
only
at discrete $\phi$-values which explain the residual values
of the conductance within the energy gap seen in
Fig. \ref{planar.fig} (b).

For the bipolar case, and for the $xy$-interface
the scattering process in the momentum space
connects points of the
FS with different sign as in the planar-wave case.
This means that the pair potential changes
sign and a ZEP is formed as seen in Fig. \ref{bipolar.fig} (a).
The tunneling spectra for the
$zy$-interface is enhanced due to the sign change caused
by the scattering $(\phi,\theta)$ to $(\pi-\phi,\theta)$
as seen in Fig. \ref{bipolar.fig} (b).

The conclusion is that the tunneling at the $zy$ interface can be used
to distinguish the pairing states with horizontal lines of nodes
on the FS.

\begin{figure}
 \centerline{\psfig{figure=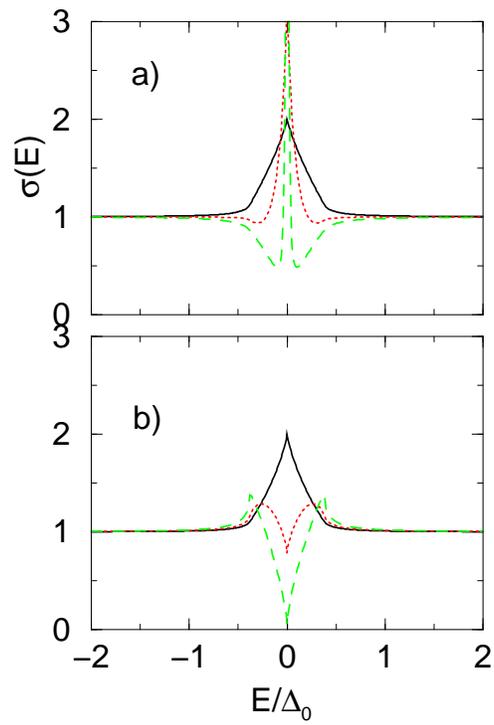,width=6.5cm,angle=0}}
  \caption{
The same as in Fig. 3.
The pairing
symmetry describes the $A$ phase of UPt$_3$ where the secondary
component of the order parameter vanishes.
}
  \label{n10.fig}
\end{figure}

In this paragraph we analyze the pairing state corresponding
to the $A$ phase (high $T$, low $H$) of UPt$_3$, where
the secondary order parameter vanishes. The resulting order
parameter  does not
break the time reversal symmetry.
For the $xy$ interface the spectra has a ZEP
due to the sign change of the order
parameter along the $z$-axis as seen in Fig. \ref{n10.fig}(a). 
Also the line shape of the spectra is
V-like because the nodes of the pair potential at $k_z=0$ intersect the FS.
For the $zy$-interface the scattering process from
$(\phi,\theta)$ to $(\pi-\phi,\theta)$ does not change the sign of the
pair potential and no ZEP occurs as seen in Fig. \ref{n10.fig}(b). However the line nodes which run parallel
to $k_z$ intersect the FS and the spectra is $V$-shaped.

\noindent{\it 5. Experimental relevance} 

In this section a comparison is made with existing tunneling experiments on 
Sr$_2$RuO$_4$ and UPt$_3$. Tunneling experiments that have been performed 
on cleaved $c$-axis junctions of Ru-embedded Sr$_2$RuO$_4$ show a 
bell shaped spectrum with a sharp peak at zero bias for the $1.4$-T 
phase and a sharp ZEP for the $3$-K phase \cite{mao}.
The spectra for the $1.4$-K phase is similar to the board ZEP 
observed in Ru-free Sr$_2$RuO$_4$ via point contact spectroscopy \cite{laube}
although their experiments actually measure the tunneling resistance.
This type of spectra for the $1.4$-K phase 
is consistent with nodeless Eu pairing state where 
the broadening of the ZEP is due to the presence of Andreev bound 
states. 
The sharp peak seen in the $3$-K phase is an indication of a pairing 
state with horizontal lines of nodes. 
It has been suggested that a phase transition occurs where 
the $3$-K phase with line node transforms to a nodeless Eu state 
close to the bulk $T_c$ \cite{sigrist}.
In the experiments of Mao {\it et al.} \cite{mao} the tunneling direction 
is along the $c$-axis. 
However the experimentalists believe that the Andreev reflection takes 
place in the in-plane direction due to the Ru inclusions. 
Then the pairing state that fits well the experimental data for the 
$zy$ interface is $d_z({\bf k})=(k_x+ik_y)\cos(ck_z)$ as seen in 
Fig. \ref{Eucosckz.fig}(b), 
rather than 
$d_z({\bf k})=(\sin(\frac{ak_x}{2})+i\sin(\frac{ak_y}{2}))\cos(\frac{ck_z}{2})$
and $d_z({\bf k})=(k_x+ik_y)^2k_z$, as seen in 
Figs. \ref{sincos.fig}(b) and \ref{Eu2kz.fig}(b).
On the other hand if the Andreev reflection occurs in the 
$c$-axis direction then the pairing state which fits more precisly the 
experimental data is $d_z({\bf k})=(k_x+ik_y)^2k_z$ which shows 
a clear ZEP for the $xy$-interface, 
for the low transparency barrier as seen in Fig. \ref{Eu2kz.fig}(a).

Point contact spectroscopy for heavy fermion superconductors UPt$_3$ has 
been performed where distinct minima in the differential resistance 
versus voltage have been observed for current flow parallel to the 
$c$-axis, and only very weak structures-if at all- have been 
observed for current flow within the basal plane \cite{goll}. 
Their observation is consistent 
with the calculated tunneling conductance for the unitary planar state 
seen in Fig. \ref{planar.fig}(a)
and \ref{planar.fig}(b), 
for the tunneling conductance along the $c$-axis and $a$-axis respectively.
Moreover measurements of the differential conductivity of 
UBe$_{13}$-Au contacts, where the UBe$_{13}$ is in polycrystalline form, 
reveal the existence of low-energy Andreev 
surface bound states, which are identified by the presence of ZEP
and are consistent with an order parameter with nontrivial symmetry 
of the energy gap \cite{walti}.

\noindent{\it 6. Conclusions} 

We calculated the tunneling conductance in 
normal-metal/insulator/triplet superconductor with horizontal lines of nodes, 
junction using the BTK formalism. We assumed 
possible pairing potentials  
for the superconductor which break the time reversal 
symmetry.
For the Sr$_2$RuO$_4$ the tunneling at the $xy$ interface can be used
to distinguish the pairing states with horizontal lines of nodes
from the ZEP that is formed when the pair potential 
changes its sign on the FS during the scattering process.
Also for the $xy$ interface 
the line shape of the spectra is V-like due to the presence 
of nodes of the pair potential along the $z$ axis 
at the FS, while for the $zy$ interface
the tunneling conductance has residual values due to the formation 
of bound states at discrete values of the angle $\phi$.

For the UPt$_3$ for the tunneling along the 
$z$-axis a ZEP is formed for the pairing states we examined, 
while the spectra along the $x$-axis has residual values or develops a ZEP 
depending on the details of the 
pairing state.
In each case the observation of a ZEP in the 
tunneling experiments
is consistent with an order parameter with nontrivial symmetry
of the energy gap.

\newpage

\end{document}